\documentclass[11pt]{article}
\usepackage{moriond,epsfig}

\def\be{\begin{equation}}
\def\ee{\end{equation}}
\def\bea{\begin{eqnarray}}
\def\eea{\end{eqnarray}}

\begin{document}
\vspace*{4cm}
\title{BiPo PROTOTYPE FOR SuperNEMO RADIOPURITY MEASUREMENTS}

\author{ M. BONGRAND\\(on behalf of the SuperNEMO collaboration) }

\address{LAL, Universit\'e Paris-Sud 11, CNRS/IN2P3, Orsay, France}

\maketitle\abstracts{
The BiPo project is dedicated to the measurement of extremely low radioactive contaminations of SuperNEMO $\beta\beta$ source foils ($^{208}$Tl~$<$~2~$\mu$Bq/kg and $^{214}$Bi~$<$~10~$\mu$Bq/kg). A modular BiPo1 prototype with its 20 modules and its shielding test facility is running in the Modane Underground Laboratory since February, 2008. The goal of this prototype is to study the backgrounds and particularly the surface contamination of plastic scintillators. After 2 months, a preliminary upper limit on the sensitivity of a 10~m$^2$ BiPo detector in $^{208}$Tl contamination of selenium source foils can be extrapolated to: $\mathcal{A}$($^{208}$Tl)~$<$~7.5~$\mu$Bq/kg (90~\% C.L.).}

\section{Principle}

 The BiPo detector is dedicated to the measurement of the high radiopurity levels in $^{214}$Bi and $^{208}$Tl of very thin materials and especially the double beta source foils of the SuperNEMO detector\cite{simard}. The expected sensitivity is $\mathcal{A}$($^{208}$Tl)~$<$~2~$\mu$Bq/kg and $\mathcal{A}$($^{214}$Bi)~$<$~10~$\mu$Bq/kg.

\begin{figure}[htb]
  \centering
  \includegraphics[scale=0.35]{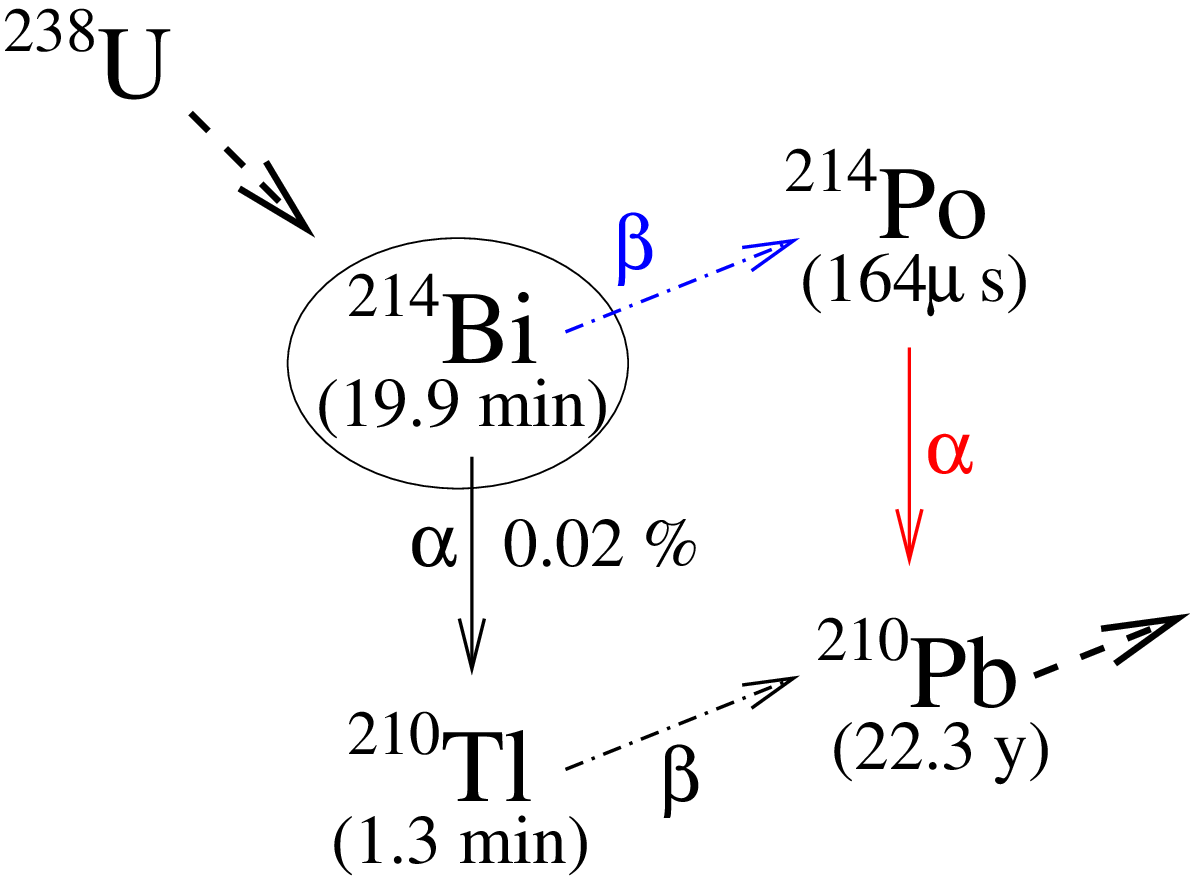}
  \hspace{1.5cm}
  \includegraphics[scale=0.35]{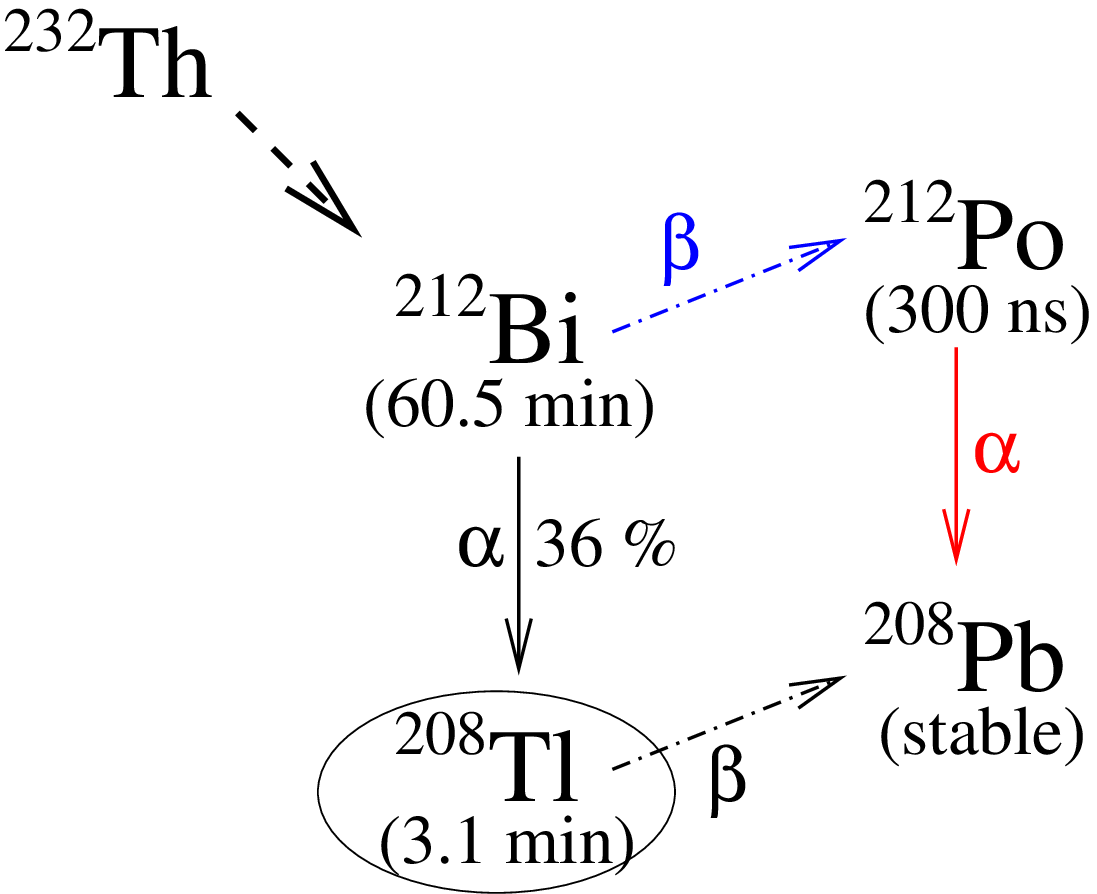}
  \caption{BiPo processes for $^{214}$Bi and $^{208}$Tl.}
  \label{fig:bipo-proc}
\end{figure}

In order to measure $^{208}$Tl and $^{214}$Bi contaminations, the original idea of the BiPo detector is to detect the so-called BiPo process, which corresponds to the detection by organic scintillators of an electron followed by a delayed alpha particle. The $^{214}$Bi isotope is nearly a pure $\beta$ emitter (Q$_{\beta}$~=~3.27~MeV) decaying into $^{214}$Po, an $\alpha$ emitter with a half-life of 164~$\mu$s (Fig.~\ref{fig:bipo-proc}). The $^{208}$Tl isotope is measured by detecting its parent the $^{212}$Bi isotope. $^{212}$Bi decays with a branching ratio of 64~\% via a $\beta$ emission towards $^{212}$Po (Q$_{\beta}$~=~2.25~MeV) which is again an $\alpha$ emitter with a short half-life of 300~ns. So, for these two chains a BiPo signature is an electron associated to a delayed $\alpha$ with a delay time depending on the isotope contamination we want to measure.

\begin{figure}[htb]
  \centering
  \includegraphics[scale=0.45]{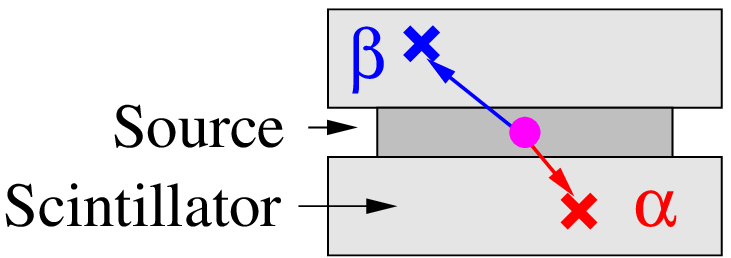}
  \hspace{1.5cm}
  \includegraphics[scale=0.35]{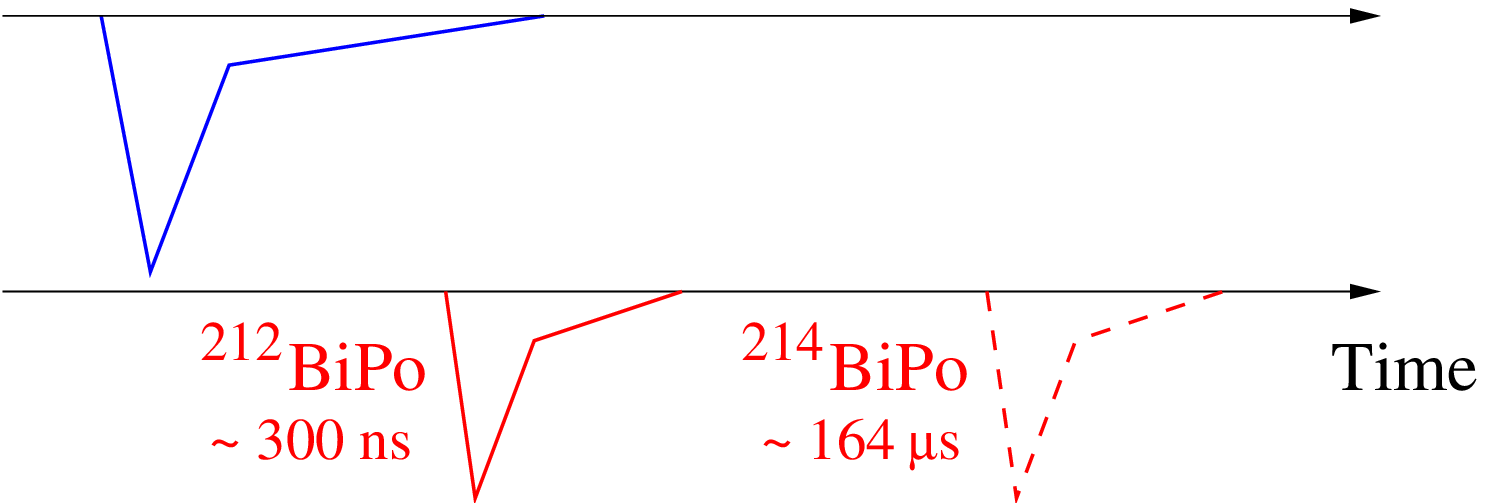}
  \caption{BiPo detection principle with plastic scintillators and time signal seen with PMTs. Dots represent the contamination and crosses represent energy depositions in scintillators (trigger in blue and delayed in red).}
  \label{fig:bipo-event}
\end{figure}

The particles emitted by the source foil are detected with plastic scintillators coupled to low radioactivity photomultipliers (Fig.~\ref{fig:bipo-event}). Plastic scintillators are very radiopure and reduce the backscattering of electrons. The detection efficiency is dominated by the capacity for an $\alpha$ particle to escape the foil. GEANT4 simulations give a total efficiency of 6.5~\% for contaminations in selenium foils (40~mg/cm$^2$) with 1~MeV threshold for $\alpha$. Therefore the energy threshold of the detector must be as low as possible. Moreover the energy converted into scintillation light is much lower for $\alpha$ compared to electrons. This quenching factor depends on the energy of the $\alpha$ and has been measured with a dedicated test bench\cite{bongrand}. For example, a 1 MeV $\alpha$ will produce same amount of light than a 40~keV electron.

\section{Backgrounds}

The BiPo measurement consists in the detection of the electron in one scintillator and the detection of the delayed $\alpha$ particle in the other scintillator. This strong BiPo signature constrains the background of the detector to only 3 processes:\\
- bismuth ($^{212}$Bi or $^{214}$Bi) contaminations in the volume of the scintillator. In such decay the electron deposits part of its energy in the first scintillator before crossing the foil to reach the other one. The delayed $\alpha$ is detectable only in the first scintillator because it can't cross the foil. This background can be rejected because two hits in time are observed in the two scintillators: this is not a BiPo event. (Fig.~\ref{fig:bipo-event}a)\\
- bismuth contaminations on the surface of the scintillator\footnote{Small thickness where the energy deposited by the electron is below the threshold ($\sim$ 100~$\mu$m for 150~keV).}. In this case the electron doesn't deposit enough energy in the first scintillator to be detected. The delayed $\alpha$ particle is still detectable only in this first scintillator. This contamination is not distinguishable from a BiPo signal because this signature exactly corresponds to a BiPo event coming from the foil. (Fig.~\ref{fig:bipo-event}b)\\
- random coincidences due to external $\gamma$. To reduce this background, the BiPo detector uses low background materials, is shielded and installed in underground lab. The single counting rate of each scintillator has to be less than 40~mHz to measure $^{208}$Tl and less than 10~mHz for $^{214}$Bi because of $^{214}$Po longer half-life. Pulse shape discrimination also reduces this background. (Fig.~\ref{fig:bipo-event}c)

\begin{figure}[htb]
  \centering
  \includegraphics[scale=0.42]{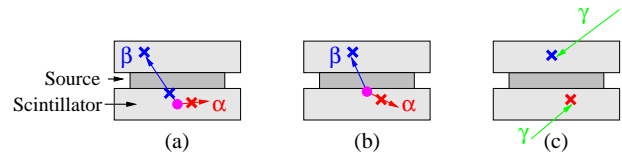}
  \caption{Surface and volume bismuth contaminations of the scintillators and random coincidences backgrounds.}
  \label{fig:bipo-background}
\end{figure}

\section{BiPo1 prototype}

\begin{figure}[htb]
  \centering
  \includegraphics[scale=0.35]{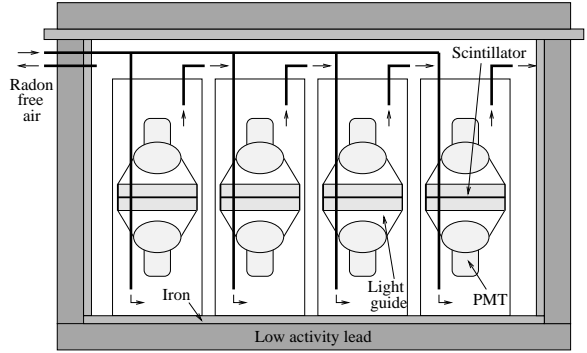}
  \caption{BiPo1 prototype in its shielding in LSM.}
  \label{fig:bipo1-setup}
\end{figure}

BiPo1 prototype is divided in 20 modules. Each module is a black box containing two polystyren based scintillators coupled to low background 5'' photomultipliers with PMMA light guides. Scintillator dimensions are 20$\times$20$\times$1 or 20$\times$20$\times$0.3~cm$^3$, the entrance window is covered with 200~nm of ultra-pure aluminum to isolate optically each scintillator and to improve the light collection. The sides of scintillators and light guides are covered with 0.2~mm of Teflon for light diffusion. The prototype is installed in the Modane Underground Laboratory (LSM) under 4800~m.w.e.. Surrounding the modules, a shielding of 15~cm of low activity lead reduces external $\gamma$ and 3~cm of pure iron stops bremsstrahlung $\gamma$ emitted in the lead by the decay of the $^{210}$Bi from long half-life $^{210}$Pb. Radon-free air flushes the volume of each module and the inner volume of the shielding (Fig.~\ref{fig:bipo1-setup}).

Photomultipliers signals are sampled with VME digitizing board\cite{breton}, during 2.5~$\mu$s with a high sampling rate (1~GS/s) and a 12~bit high dynamic range (1~V). The acquisition is triggered each time a pulse reaches the 150~keV energy threshold and the 2 photomultipliers signals from a module are stored. The delayed hit research is performed later by the analysis of the signals (Fig.~\ref{fig:bipo1-c1-results}).

\section{BiPo1 calibration}

\begin{figure}[htb]
  \centering
  \includegraphics[scale=0.34]{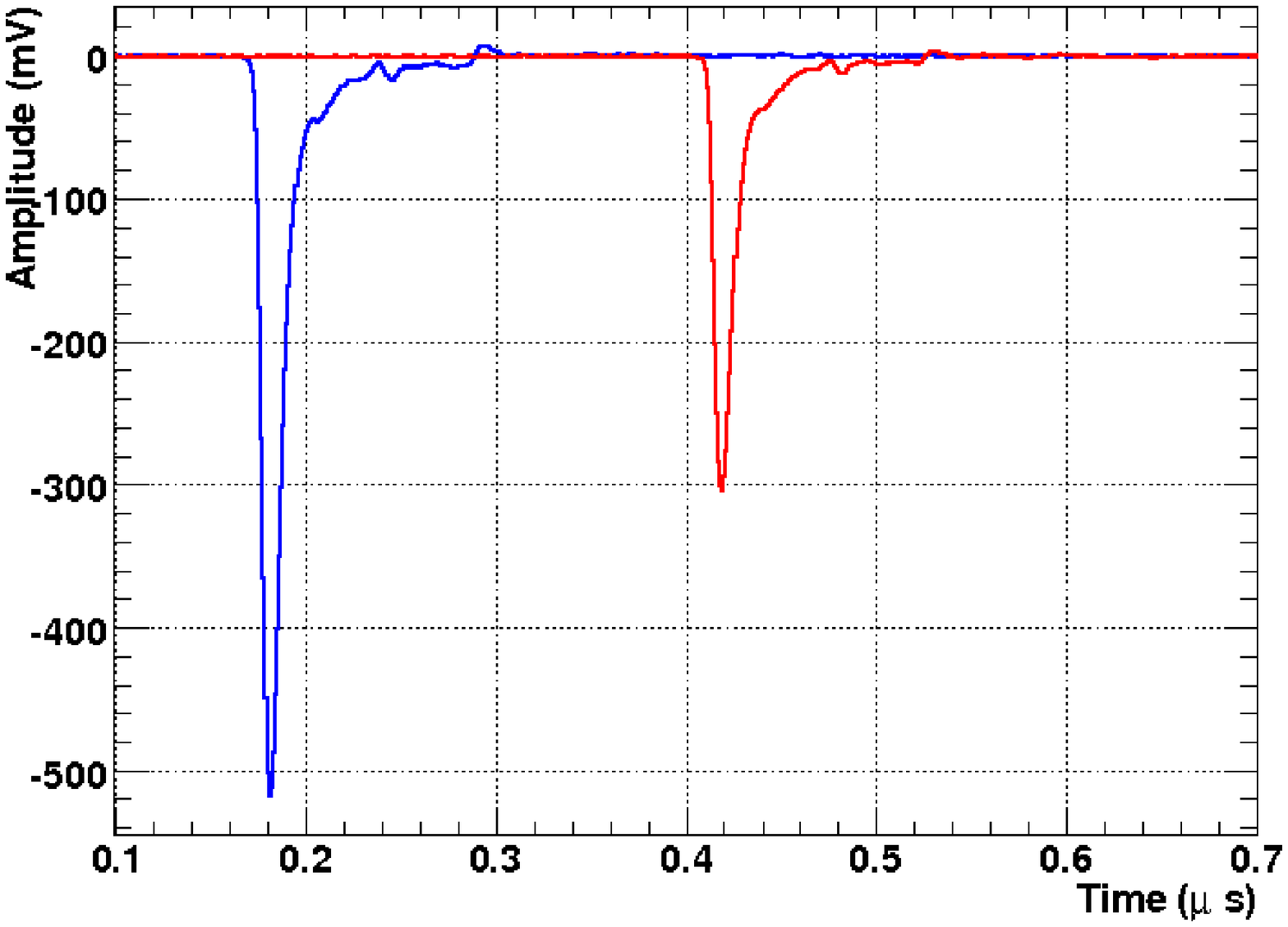}
  \includegraphics[scale=0.34]{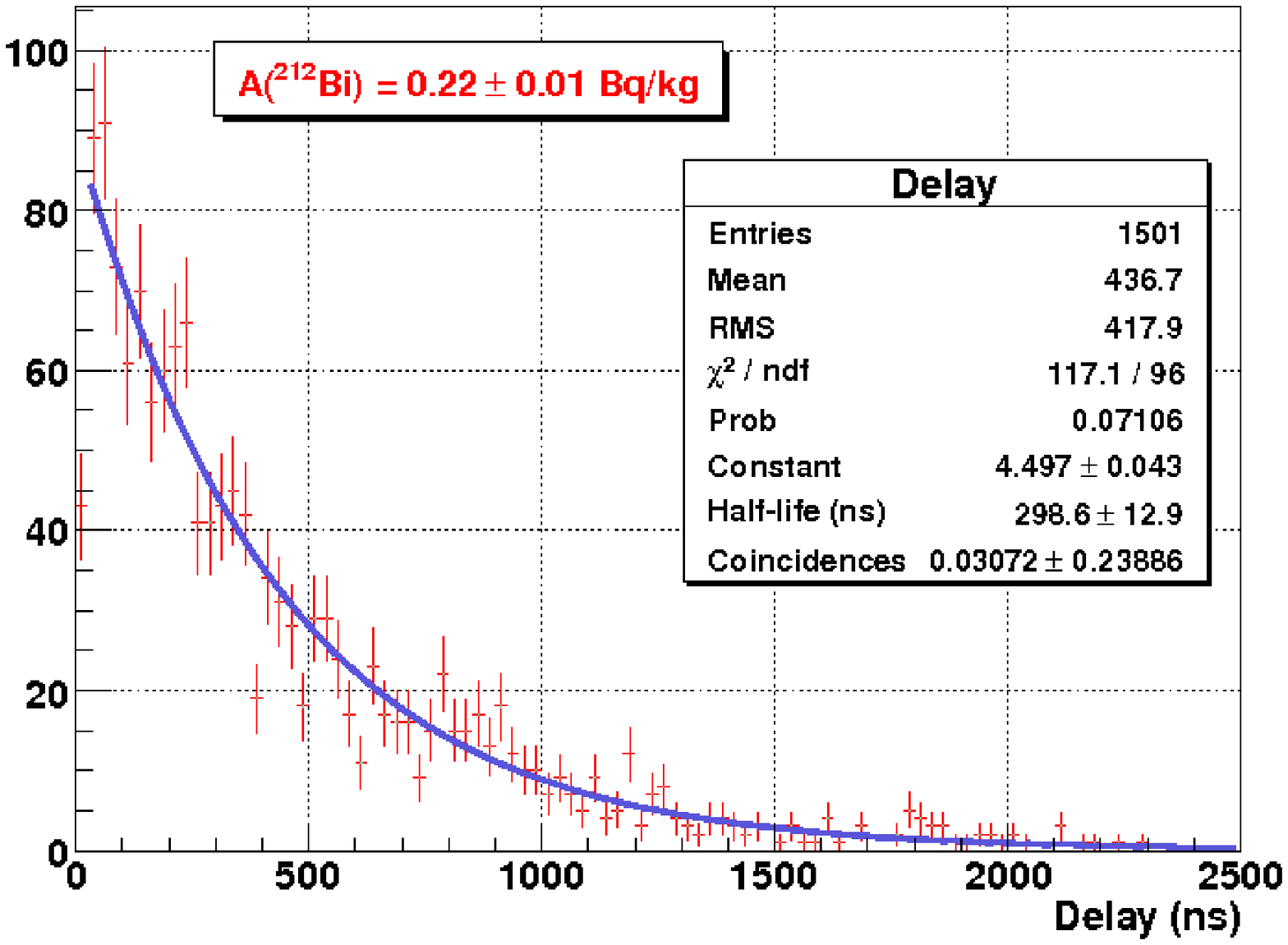}
  \caption{Example of a BiPo event observed in BiPo1 and delay distribution between the $\beta$ and the $\alpha$ decays.}
  \label{fig:bipo1-c1-results}
\end{figure}

The first BiPo1 module has been dedicated to the validation of the detection principle for $^{212}$Bi contaminations inside a foil. A 150~$\mu$m aluminum foil (40~mg/cm$^2$) with a contamination measured with HPGe detectors of $\mathcal{A}$($^{212}$Bi$\rightarrow ^{212}$Po) = 0.19 $\pm$ 0.03 Bq/kg, has been installed between the two scintillators. After 141~days of data taking, 1501 BiPo events have been detected. Taking into account the efficiency calculated by GEANT4 simulations, it corresponds to a reconstructed activity of $\mathcal{A}$($^{212}$Bi$\rightarrow ^{212}$Po) = 0.22 $\pm$ 0.01 Bq/kg, in good agreement with initial HPGe measurement. The delay between the two hits is also measured, and the fit of the decay law perfectly corresponds to the $^{212}$Po half-life (Fig.~\ref{fig:bipo1-c1-results}). These results are a strong validation of the measurement principle and the calculated efficiency.

\begin{figure}[htb]
  \centering
  \includegraphics[scale=0.34]{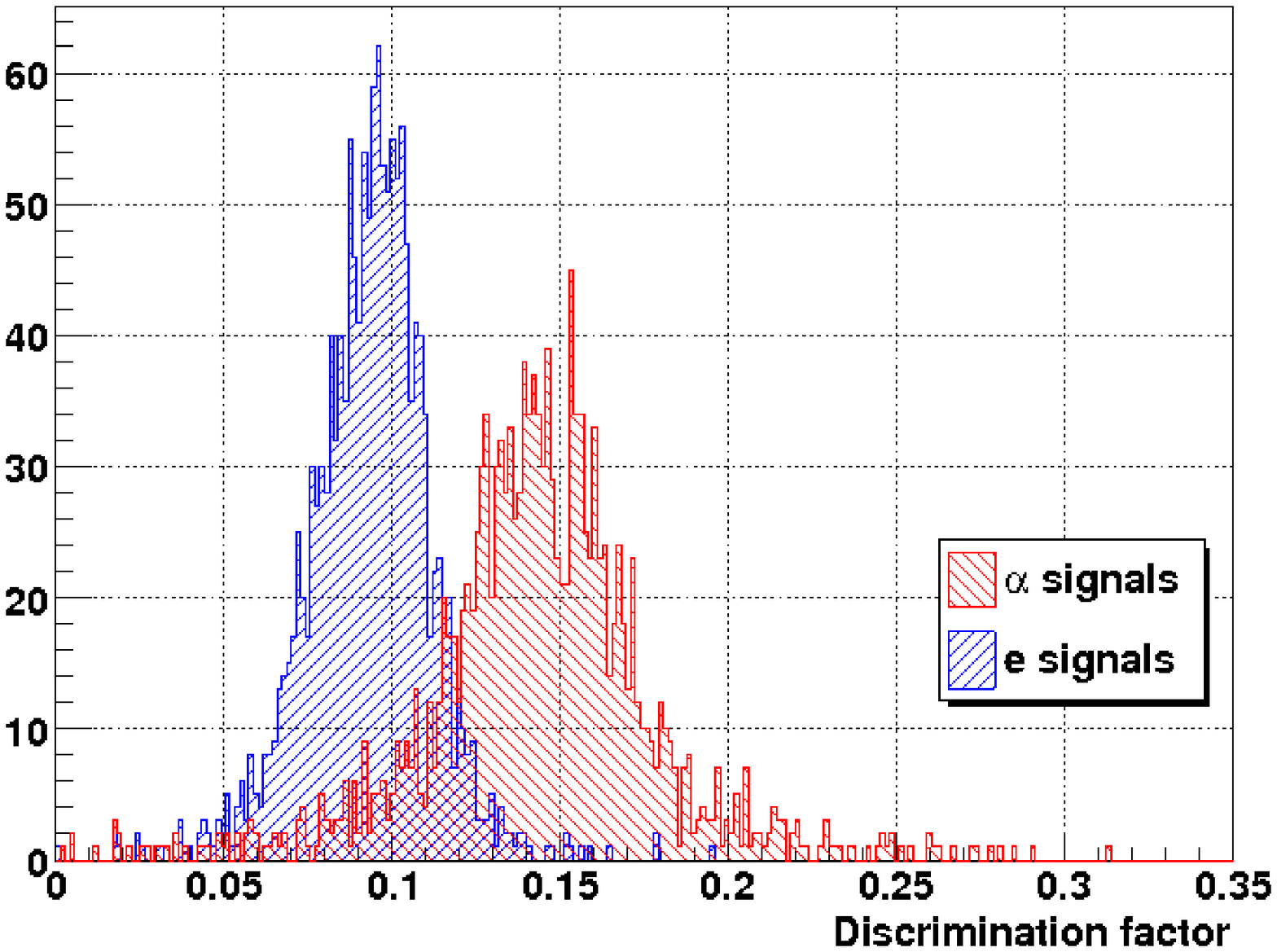}
  \includegraphics[scale=0.34]{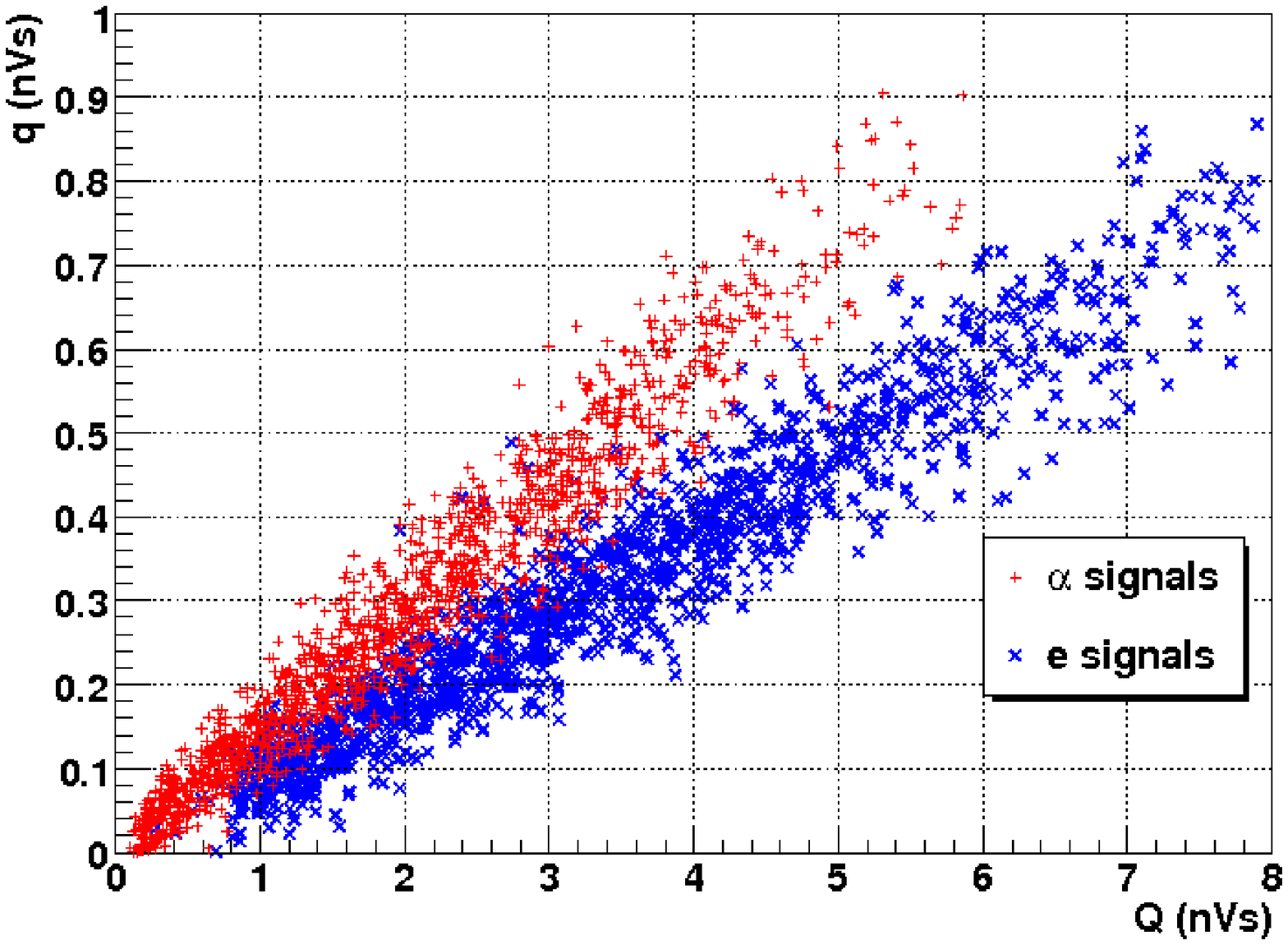}
  \caption{Distributions of the tail-to-total charge ratio and the tail charge $q$ as a function of the total charge $Q$ for the prompt and delayed signal from the aluminum foil in the first module of BiPo1.}
  \label{fig:bipo1-c1-discri}
\end{figure}

Particle identification in the plastic scintillator could help to sign the BiPo process and reject random coincidences of $\gamma$. Indeed, longer half-life states in scintillators are excited by $\alpha$ particles but not by electrons. More light is therefore observable in the tail of the pulse for $\alpha$ particles. A pulse shape discrimination, using tail-to-total charge ratio as discrimination factor, has been applied on the photomultiplier signals from the data of the calibration foil. A good separation has been observed for prompt ($e^-$) and delayed ($\alpha$) signals (Fig.~\ref{fig:bipo1-c1-discri}). Using this discrimination, it is possible to reject 80~\% of the random coincidence background and to keep 90~\% of the true BiPo events.

\section{BiPo1 surface radiopurity}

The other modules of BiPo1 are dedicated to the measurement of the surface radiopurity of scintillators. BiPo events, coming from the contact surface between the two scintillators, are observed by a hit in one scintillator and a delayed hit in the other one. After 2 months of data taking, 7 BiPo events have been observed on a statistics equivalent to 0.8~m$^2 \times$month. It corresponds to an activity of about 2~$\mu$Bq/m$^2$. Extrapolating this background to a final 10~m$^2$ BiPo detector for a one month measurement of 4~kg selenium source foil for SuperNEMO, the thallium preliminary sensitivity is: $\mathcal{A}$($^{208}$Tl)~$<$~7.5~$\mu$Bq/kg \small(90\% C.L.)\normalsize.

\section{Conclusion}

The BiPo1 prototype demonstrated the validity of the experimental technique to measure $^{208}$Tl contaminations in thin materials. The preliminary sensitivity achieved is 10 times better than standard HPGe measurements. Particles identification with simple plastic scintillators enhance the performances of the prototype. New modules using ``phoswich'' scintillators, a compound of a thin fast scintillator to detect $\alpha$ and a thick slow scintillator for electrons, will improve the discrimination between the 2 particles. A second prototype, BiPo2, using large scintillator plates (0.56~m$^2$) will also be tested in LSM before this summer.

\end{document}